\documentstyle[array,preprint,tighten,aps]{revtex}
\begin{document}

\title{Growth in systems of vesicles and membranes}
\author{ A.M. Somoza$^{a,c}$, U. Marini Bettolo Marconi$^b$ and P. Tarazona$^{c}$ }
\address{$^{a}$ Instituto de Ciencia de Materiales, Consejo Superior 
de Investigaciones Cient\'{\i}ficas, Madrid-28049, Spain \\
$^{b}$ Dipartimento di Matematica e Fisica, Universita degli
Studi di Camerino, via Madonna delle Carceri, I-62032 Camerino, Italy \\
$^{c}$ Departamento de F\'{\i}sica de la Materia Condensada and Instituto
Nicolas Cabrera, Universidad
Aut\'onoma de Madrid, Madrid E-28049, Spain}

\date{\today}
\maketitle

\begin{abstract}
 We present a theoretical study for the intermediate stages of the growth of
membranes and vesicles in supersaturated solutions of amphiphilic
molecules.  The problem presents important differences with the growth
of droplets in the classical theory of Lifshitz-Slyozov-Wagner,
because the aggregates are extensive only in two
dimensions, but still grow in a three dimensional bath. The balance 
between curvature and edge energy favours the nucleation 
of small planar membranes, but as they grow  beyond a critical size 
they close themselves to form vesicles. 
We obtain a system of coupled equations describing the growth of
planar membranes and vesicles, which is solved
numerically for different initial conditions. Finally, the range of
parameters relevant in experimental situations is discussed.
\end{abstract}

\section{Introduction}

Solutions of amphiphilic molecules in water may form a large variety
of molecular aggregates, as a result of the asymmetric interaction 
with water of the hydrophilic heads and the hydrophobic tails 
of these molecules
\cite{Lindman85,Seddon90,Gelbart94,Gompper94,Israelachvili85}.
These aggregates range from micelles, which may be considered
as large molecular clusters, to continuous structures of the amphiphile
with regular or irregular structures such as sponge phases. 
In between, one finds
structures like membranes and vesicles which are macroscopic
in two dimensions, whereas they are only a few molecular 
sizes thick. From a thermodynamic point of view one can regard
membranes as two dimensional phases in
coexistence with a diluted bulk solution of water and 
amphiphile molecules,
but contrary to the case of adsorbed layers on solid
substrate, the membrane is not restricted to lie on fixed
positions and orientations. The membrane is
free to explore the full three dimensional space,
and in the case of ``fluid membranes'',
the lack of rigid molecular order within the membrane surface
allows the bending of the membranes, with low energetic cost
as ``curvature energy'' \cite{Helfrich73}. The extraordinary properties
of membranes as self-assembling, self-sealing, insulating and flexible
structures
are used in biological systems to form the basis of cellular membranes,
plasts and mitocondrial structures, the insulating material
for neurons, etc. In the last decade, the study of membranes and
other amphiphilic aggregates has attracted the attention of
experimental and theoretical physicists, 
chemists, physical-chemists, biologists 
\cite{Nelson89,Lipowski91,Lipowski92,biomembranes}. 

  In the present paper we discuss the dynamics of formation
of these aggregates from a supersaturated bulk solution,
by extending to the peculiarities of these systems the methods developed for
``regular'' three dimensional phases. 
 We focus our interest on amphiphilic systems forming  two types of 
isolated
bilayer aggregates: membranes and vesicles.
During the formation of these macroscopic
aggregates we may distinguish two different stages:\\
{\it (i)} A {\it nucleation} process, during which the
microscopic aggregates form.\\
{\it (ii)} A {\it growth} process,
when these seeds reach mesoscopic sizes or dissolve into the
bulk solution, by following a nearly deterministic dynamics
which transforms the initial population of the microscopic
clusters into a distribution of sizes for the aggregates
and depletes the amphiphile concentration in bulk solution 
to the equilibrium value.\\
 In an infinite system at zero temperature, the growth process would
never end. Alternatively, we should consider a third regime when the
super-saturation is so small that the gradients in the chemical
potential are affected by fluctuations.

 Experimentally, these aggregates can be obtained with the help 
of techniques designed to
accelerate the nucleation process.
 Among these are the ultrasound technique and the
use of solid substrates (heterogeneous nucleation), which preferentially 
adsorb the amphiphilic molecules.
On the other hand, the spontaneous formation from 
an equilibrium super-saturated solution is of interest. In fact, 
it has been suggested \cite{Gene} that, due to the possibility of
spontaneous formation, amphiphilic systems forming vesicles played
a crucial role in the beginning of life, providing for the 
compartmentation of the early metabolic and self-reproducing
molecular machinery.

 In this work we address the growth process, in which the
aggregates formed in the early stage compete with each other
to grow, incorporating the amphiphilic molecules from the
super-saturated solution. We try to answer the questions
about how the typical distribution of sizes evolves and how it depends
on the distribution in the early stage.  Our model needs
as input the initial distribution of sizes as well as the initial
super-saturation of the system. This data could be obtained from
nucleation theory, but
the study of the different mechanisms of nucleation requires
molecular models of these complex systems, which are still
in an early stage of development. 

\section{The model}

Let us consider the possible equilibrium shapes for bilayer 
aggregates. For a fixed  area, $A$, a symmetric bilayer adopts the shape that
minimizes the effective Helfrich Hamiltonian \cite{Helfrich73}:
\begin{equation}
{\cal H} = \int dA \left\{ \frac {1}{2} \kappa (C_1 + C_2)^2 + \kappa_G C_1
C_2 \right\} + \lambda \oint dl,
\label{Hamil}
\end{equation}
where the first integral extends over the whole surface area and the other
one extends along the boundary, where $\kappa$ and  $\kappa_G$ are
respectively the bending 
rigidity  and the Gaussian curvature moduli, $\lambda$ is the
line tension and $C_1, C_2$ are the two principal local curvatures
of the surface. In the case of a flat circular membrane 
of radius $R$, ${\cal
H} \propto R$, whereas for a
spherical vesicle of radius $S$, ${\cal H} = 4 \pi ( 2\kappa
+\kappa_g)$. Thus, small aggregates tend to assume a planar shape but larger
aggregates will always prefer a closed shape. To locate the
spontaneous shape transformation we employ a simple
{\it spherical cap model} where the bilayer is restricted to adopt the
shape of the curved surface of a sphere cut by a plane. For a fixed area
the shape can be fully described by 
the radius, $R$, of the circle of intersection between the sphere and the
plane,
\begin{equation}
{\cal H}(A,R)= 4 \pi \kappa_s \left( 1 - \frac{\pi R^2} {A}\right ) +2 \pi
\lambda R,
\end{equation}
where $\kappa_s=2 \kappa + \kappa_G$. A spherical vesicle
corresponds to $R=0$ while a planar membrane is described by 
$R=(A/\pi)^{1/2}$. 
In Figure 1 we display the energy $\cal H$ versus $R$ for different values of
$A$. The absolute minimum is always at an extremum: for 
$A< 4 \pi (\kappa_s/\lambda)^{2}$ it corresponds to
planar membranes while for $A > 4 \pi (\kappa_s/\lambda)^{2}$ it corresponds
to spherical vesicles. When the two configurations have the same energy
($A= 4 \pi (\kappa_s/\lambda)^{2}$), they are separated by an
energy barrier $\Delta {\cal H}= 3 \pi \kappa_s$, which decreases
with increasing $A$ and finally disappears at $A_c=\pi(4\kappa_s/\lambda)^{2}$,
where all planar membranes become unstable and close into vesicles. 

 Such a simple model gives a first glimpse of the 
growth process: unless the nucleation mechanism is strongly
biased to produce vesicles, the early population of aggregates
consist of planar membranes; as some of these grow they 
become metastable with respect to the closed vesicles, but still
separated from them by a barrier, until their area reaches the critical 
value $A_c$. At this stage the membranes become unstable
and turn into closed vesicles of the same
area, which can grow until the solution has been depleted
down to the equilibrium concentration of amphiphiles. 

 We start constructing our model 
of growth for a system of
both vesicles and membranes by making the following assumptions:\\
{\it (i)} The excess concentration (with respect to a dilute solution
without aggregates) is so small that the interaction between aggregates
can be neglected.\\
{\it (ii)} All aggregates are much larger than their typical thickness ($
\xi \approx 40 \AA$). This assumption is expected to
hold at least in the long 
time regime.\\
{\it (iii)} The main mechanism for transport of
amphiphiles is diffusion as expected when the water
solution is at rest, so that all hydrodynamic effects can be neglected.
In this sense it is implied that the growth process does not agitate the water
solution.\\
{\it (iv)} We consider an infinite volume.\\
{\it (v)} Finally we assume that fluctuations can
be neglected.

We shall later comment on some of these assumptions.
Within this approximations the problem clearly resembles the
classical Lifshitz-Slyozov-Wagner (LSW)  theory for the growth of
spherical domains \cite{LS,Wagner}. 

Within a quasi-equilibrium thermodynamic description the chemical 
potential, $\mu({\bf r})$,
changes smoothly with 
position. We consider an isolated growing aggregate of area $A$ and $N(A)$ 
molecules;
its rate of growth will depend on the number of amphiphilic
molecules that approach its surface:
\begin{equation}
   {{ d N(A)} \over { d t}} = - \oint {\bf j} \cdot d{\bf S}.
\label{dNdt}
\end{equation}
where the integral extends over a surface enclosing the aggregate
and ${\bf j}({\bf r})$ is the current of amphiphilic molecules induced
by the gradients of chemical potential which is different in the bulk 
and at the surface of
the aggregate:
\begin{equation}
{\bf j}= -\alpha \nabla \mu({\bf r}),
\label{current}
\end{equation}
$\alpha$ being a kinetic coefficient. 
 Assuming that the transport of amphiphiles does not change
its local concentration,
\begin{equation} 
\nabla \cdot {\bf j}=0,
\label{poisson}
\end{equation}
everywhere outside the border of the aggregate. This leads to a Poisson
equation for $\mu({\bf r})$, equivalent to an electrostatic potential with
boundary conditions at infinity, $\mu_{\infty}$, and at the 
surface of the aggregate $\mu(A)$, which mimics an equipotential
metallic boundary. Such a value, $\mu(A)$, depends on the particular 
properties of each aggregate.

\subsection{Membranes.}

Let us consider a system composed only of membranes and assume that none
of them transforms into vesicles.
The excess of grand potential energy of a planar circular membrane of 
radius $R$ is
\begin{equation}
\Delta \Omega_m(R) = \pi \sigma R^2+ 2 \pi \lambda R,
\label{Omega_m}
\end{equation}
where $\sigma$ is the surface tension of the membrane and $\lambda$ is the
line tension associated with the boundaries.
The equilibrium of a large membrane of arbitrary shape is 
controlled by
the requirement $\sigma=0$, which at a given temperature is satisfied 
when the chemical potential $\mu$ assumes its equilibrium 
value $\mu_o$. In general, for a
fixed chemical potential $\mu$, the surface tension does not vanish
but is given by:
\begin{equation}
\sigma(\mu)= - \Gamma (\mu-\mu_0) 
\label{GD}
\end{equation}
where $\Gamma$ is the adsorption per unit area in the membrane. Since we
are
considering systems with small supersaturations, within a 
linear approximation we take the values of $\lambda$ and $\Gamma$ 
evaluated at $\mu=\mu_o$. For
$\mu > \mu_0$ eq. (\ref{Omega_m}) has a maximum at
\begin{equation}
R_c(\mu)=\frac {\lambda} {\Gamma (\mu - \mu_0)}.
\end{equation}
In other words, membranes with $R>R_c(\mu)$ tend to adsorb particles
and grow. This growth is much faster
than the diffusion process described above (provided there are particles 
to be absorbed in the neighbourhood of the membrane), and it will 
stop only when the adsorption of particles has effectively changed the chemical
potential in the proximity of the membrane. We thus obtain the boundary
condition of the diffusion problem:
\begin{equation}
\mu(R)=\mu_0 + \frac {\lambda}{\Gamma R}.
\end{equation}
 An equivalent argument applies for shrinking membranes when $\mu< \mu_0$.
Now it is possible to get the chemical potential field for an isolated
membrane at an arbitrary distance $\bf r$ from its center. Neglecting 
the thickness of the membrane is equivalent to know the
electrostatic potential created by a planar metallic disk \cite{Arfken}:
\begin{equation}
\mu({\bf r})=\mu_{\infty}+\frac{2(\mu(R)-\mu_{\infty})R}{\pi r}
\sum_{l=0}^{\infty} \frac{(-1)^l}{2l+1} \left( \frac{R}{ r}\right)^{2l} 
P_{2l}(\cos \theta)
\label{disk}
\end{equation}
for $R<r$, where ${\bf r}=(r,\theta,\phi)$ in spherical coordinates.
From eq. (\ref{current}) and (\ref{dNdt}), the rate of growth of an isolated
membrane of radius $R$ is
\begin{equation}
\frac {dR}{dt}=\frac {4 \alpha \lambda} {\pi \Gamma^2}
\left(\frac {\Gamma \Delta}{\lambda} -\frac{1}{R}\right),
\label{dRdt}
\end{equation}
where $\Delta=\mu_{\infty}-\mu_0$ is a measure of the 
supersaturation of the system.
 
 We turn now to the study of an ensemble of
{\it isolated} membranes by introducing
a size distribution function, $f_m(R,t)$ which gives the number of membranes
of radius $R$ per unit volume at time $t$. $f_m(R,t)$ verifies the continuity
equation
\begin{equation}
\frac{\partial f_m}{\partial t}+\frac{\partial}{\partial R}
\left(f_m v_R\right)=0,
\label{continuity}
\end{equation}
where $v_R=dR/dt$ given by eq. (\ref{dRdt}).
 Finally, a closed set of equations is obtained by imposing
the conservation of the total amount of amphiphiles:
\begin{equation}
\chi \Delta(t)+\int_0^{\infty} \hspace{-0.2cm}dR 
\hspace{0.1cm} \pi R^2 \Gamma f_m(R,t)=Q,
\label{conservation}
\end{equation}
where the first term represents the excess number of particles per unit volume 
that remain in solution ($\chi$ is the bulk compressibility) and the integral
gives the number of particles per unit volume that belong to the 
membranes.
With initial conditions for $f_m(R,t)$ and $\Delta(t)$ the system of
equations
(\ref{dRdt}), (\ref{continuity}) and (\ref{conservation}) is
fully determined.
Following the ideas of the LSW theory it is straight-forward to get 
the asymptotic behaviour of the system (see for example \cite{LP}). 
We summarize the main results:
\begin{equation}
\Delta(t)=\left( \frac {\pi \lambda}{2 \alpha t}\right)^{1/2},
\end{equation}
\begin{equation}
N(t)=\frac {Q \Gamma}{\alpha \lambda I t},
\end{equation}
\begin{equation}
{\bar R(t)}=R_c(t)=\left(\frac{2\alpha\lambda t}{\pi\Gamma^2}\right)^{1/2},
\label{R_c}
\end{equation}
\begin{equation}
{f_m(R,t)}=\frac{N(t)}{R_c(t)} \ P\Bigl(\frac{R}{R_c(t)}\Bigr),
\end{equation}
where $N(t)$ is the total number of membranes per unit volume, $\bar R(t)$ is
the mean radius and $P(x)$ and
$I$ are 
\begin{equation}
P(x)= \left\{
 \begin{array}{ll}
\frac{8x}{(2-x)^4} \exp\left(-\frac{2x}{2-x}\right) & 
\mbox{ \ \ \ \ } 0 \leq x \leq 2 \\
& \\
0                                 & \mbox{ \ \ \ \ } 2 < x,
\end{array}
\right.
\end{equation}
\begin{equation}
I=\int_0^{\infty} dx \ x^2 P(x) =1.1094 
\end{equation}
It is important to note that within the present model the 
growth rate $R \propto t^{1/2}$
with a growth exponent $n=1/2$ instead of 1/3 as predicted by LSW.
 Such a difference is due to the two-dimensional nature of 
the aggregate, which allows for a faster growth.

Our asymptotic formulae are expected to be valid for 
\begin{equation}
log^2 \left( \frac {4 \alpha \Delta^2_0}{\pi \lambda} t \right) \gg 1,
\label{log}
\end{equation}
where $\Delta_0$ is the initial supersaturation of the system.
Unfortunately this equation does not always hold because before reaching 
a long enough time some membranes can transform into vesicles. Thus, this
restriction must hold before $R_c(t) \leq 2 R_T \approx 8 k_s /\lambda$.
From eq. (\ref{R_c}), eliminating $t$ in (\ref{log})
\begin{equation}
log^2 \left( \frac {8 \Gamma \Delta^2_0 k_s}{\lambda^4}\right) \gg 1.
\label{log2}
\end{equation}
Taking rough estimates of the parameters shown in Table I we get $18$
for the left-hand-side of eq. (ref{log2}). Thus,
real systems hardly reach this universal distribution function.

\subsection{Vesicles.}

We now consider the case of a system formed only by spherical vesicles and no
planar membranes.
In the case of an isolated vesicle of 
radius $S$, the excess grand potential is:
\begin{equation}
\Delta \Omega_v= -4 \pi \Gamma S^2 (\mu-\mu_0) + 4 \pi \kappa_s.
\label{omegav}
\end{equation}
 It is clear from this equation that all vesicles, independently of their
size, will grow until the chemical potential in the surroundings is $\mu_0$.
Thus, it is easy to get the three equations that define the problem:
\begin{equation}
\frac{dS}{dt}=\frac {\alpha \Delta} {2 \Gamma},
\label{dSdt}
\end{equation}
\begin{equation}
\frac{\partial f_v}{\partial t}+\frac{\partial}{\partial S}\left (f_v v_S\right)=0,
\label{contivesi}
\end{equation}
and
\begin{equation}
\chi \Delta(t)+\int_0^{\infty} \hspace{-0.2cm}dS \hspace{0.1cm}4\pi S^2
\Gamma f_v(S,t)=Q,
\label{conservesi}
\end{equation}
This set of equations lends itself to
an analytical solution in terms of the initial conditions. The main
result is that the size distribution function does not change its shape,
\begin{equation}
f_v(S,t)=f_0\Bigl(S- \frac{\alpha}{2\Gamma}\int_0^t dt' \Delta(t')\Bigr)
\label{fst}
\end{equation}
where $f_0(S)$ represents
 the initial distribution of vesicles. In this case $\Delta(t)$
decays exponentially. The reason for this behaviour is clear from eq. 
(\ref{dSdt}), which indicates that all vesicles grow at the same rate
independently of their size.
 
  It is interesting to notice that eqs. (\ref{dSdt}-\ref{conservesi}) were
obtained under the assumption that the slowest growth process is the
diffusion of particles (diffusion limited growth). But, for a vesicle to
grow, it is necessary to fill the interior with water, thus a certain amount
of water needs to overcome an energy barrier when crossing the bilayer
structure. 
Under some circumstances this can be the slowest process, and then 
eq. (\ref{dSdt})
should be modified. 
In this case the number of water molecules that cross the bilayer per 
unit time will be proportional to the area of the vesicle. For small enough
supersaturations (or large enough curvature constant) the shape of the 
vesicle will remain spherical, and the change in volume is :
\begin{equation}
\frac {dV}{dt}=4\pi S^2 \frac{dS}{dt}= 4\pi S^2 F(\Delta)\approx C S^2 \Delta
\end{equation}
where $F(\Delta)$ is a function related to the properties of the barrier
that verifies $F(0)=0$ and, thus, it is expected to show a linear behaviour
for small
supersaturations, $C$ being a constant. This equation is equivalent to
eq. (\ref{dSdt}) except for a constant, thus the basic result, eq. (\ref{fst})
still holds except for the time scale.

\subsection{Membranes and Vesicles.}

When there are no more membranes closing themselves to form vesicles,
the distribution function $f_v$
does not change in shape, it merely moves to larger sizes until
all the excess particles in the solution are exhausted. But in order to
relate the distribution function when the nucleation process finishes
it is necessary to allow for the existence of both membranes and vesicles and
permit the spontaneous transformation from one to the other. We study this 
case assuming that this transformation occurs at one particular radius,
$R_T=4 \kappa_s/\lambda$ and $S_T=R_T/2$, 
(estimated from the spherical cap model) and the time scale of the process
is totally negligible compared with diffusion times. Now we have to consider
size distribution functions for both membranes and vesicles.

\begin{equation}
\frac{\partial f_m} {\partial t}+\frac{\partial}{\partial R}\left(
v_R f_m\right)=-\delta(R-R_T) I_m
\end{equation}

\begin{equation}
\frac{\partial f_v} {\partial t}+\frac{\partial}{\partial S}\left(
v_S f_v\right)=\delta(S-S_T) I_m
\end{equation}

\begin{equation}
\chi \Delta(t)+
\int_0^{\infty} \hspace{-0.2cm}dR \hspace{0.1cm}\pi R^2 \Gamma f_m(R,t)+
\int_0^{\infty} \hspace{-0.2cm}dS \hspace{0.1cm}4\pi S^2 \Gamma f_v(S,t)=Q,
\label{consertot}
\end{equation}
where $I_m$ is the number of membranes per unit volume that transform into
vesicles at time $t$, $I_m=max(0, v_{R_T} f_m(R_T,t)$.

We can change to dimensionless units:
\begin{equation}
\begin{array}{ll}
R=x L, S=x L/2,     &      L=\frac{\lambda \chi}{\Gamma Q},           \\
t=\tau T,           &      T=\frac{\pi\lambda\chi^2}{4\alpha Q^2},    \\
\Delta= \delta D,   &      D=\frac{Q}{\chi},                          \\
f_m=f_v=\varphi F,\mbox{\hspace{2cm}}  &      F=\frac{\Gamma^2Q^4}{\pi\lambda^3\chi^3},  \\
\end{array}
\label{units}
\end{equation}
where we have made use of the fact that there are no membranes for $R>R_T$
and there are no vesicles for $S<S_T$ to join both size distribution functions
into one single function $\varphi(x,\tau)$: for $x<x_T\equiv R_T/L$ it
represents membranes of radius $R=x L$ and for $x>x_T$ vesicles of radius 
$S=x L/2$ . With these changes of variable
the new equations read:
\begin{equation}
\begin{array}{ll}
\frac{\partial\varphi}{\partial\tau}+\frac{\partial}{\partial x}\left(
(\delta -\frac{1}{x})\varphi\right)=0&  \ \ \   for \ \ \ \ x< x_T\\
                   &                                 \\
\frac{\partial\varphi}{\partial\tau}+\frac{\partial}{\partial x}\left(
\frac{\pi}{4}\delta\varphi\right)=0&  \ \ \   for \ \ \ \ x> x_T,
\end{array}
\end{equation}
with a special boundary condition at $x=x_T$, and
\begin{equation}
\delta(\tau)+\int_0^{\infty}\hspace{-0.2cm}dx\hspace{0.1cm}x^2\varphi(x,\tau)=1
\end{equation}

These equations have been solved numerically. We have applied a standard 
{\it upwind} algorithm \cite{numrecipes} discretizing both in time
and in space with $\Delta\tau=0.02$ and $\Delta x=0.1$. 
 The initial conditions were chosen having in mind the particular values of
the parameters shown in Table I. We consider that initially we only have
membranes and assume that the size distribution function is 
gaussian with mean ${\bar R_i} = 10 \psi \approx 4 \times 10^{-8} m$ and a
width $\sigma_i={\bar R_i}/4=10^{-8} m$. Our study applies when the 
nucleation process is already completed. The energy barrier for the 
formation of new membranes must be much lower that $k_B T$; this means that
$\Delta_0\ll \pi\lambda^2/(\Gamma k_BT)\approx 8\times 10^{-21}J/molec$.
We have selected $\Delta_0=3\times 10^{-23} J/molec$ which corresponds to 
a critical radius $R_c(0)=3{\bar R_i}/4$. The only parameter that remains
to be fixed is the height of the initial distribution, or equivalently the
value of $Q$. We arbitrarily consider that at the initial time $90\%$ of the
total excess of particles is already in a membrane 
whereas only  the remaining $10\%$ is
solved in water, leading to $Q=8\times 10^{16} molec/m^3$. The results of
the calculation are shown in Figure 2. Instead of the size distribution
function (whose total integral tends to zero in time)  we plot
$x^2 \varphi(x,\tau)$ which is proportional to the probability of finding
a molecule in an aggregate of size $x$ at time $\tau$, and its total integral
tends to one. In the early stage the distribution function moves to
larger sizes and spreads, decreasing the height of the peak.  At $\tau \approx
1300$ some membranes start to transform into vesicles and the rate increases
until $\tau \approx 3000$ when it starts to decrease. At $\tau\approx 5200$
the critical radius becomes larger than the transformation radius and ,thus
all membranes shrink. At that time the final distribution function for
vesicles is known; it will just translate to larger sizes absorbing all
remaining particles in the solution and in the membranes. In Figure 3 
is shown the critical radius, $x_c$ versus $\tau$. For these initial conditions
the behaviour is quite monotonous. I is almost linear from $\tau=500$ to
$\tau =6000$. For larger times, when most of the membranes have disappeared 
we would reach the expected exponential behaviour.

 A totally different behaviour can be obtained from different initial
conditions. The results are shown in Figures 4 and 5. The only difference with
respect to the previous case is that we have assumed that at the initial
time only $5\%$ (instead of $90\%$) of the total excess of particles is forming
the initial membranes. In the early time the critical radius is lower than the
mean radius and there is a large amount of material 
to be absorbed; this permits a 
very fast growth of the membranes without broadening the distribution function
and as soon as $\tau \approx 100$ there are already some vesicles formed
(this early time growth process is not shown in the figure because it is
covered by later data). This fast growth also makes the critical radius 
increase and, as soon as some membranes are formed, it crosses the peak of the
distribution of membranes. As the distribution function is highly peaked
this crossing suddenly makes most of the membranes shrink. For a
period of time there is an equilibrium where the critical radius remains
constant as the vesicles grow at the expense of the membranes. When the
number of membranes is small the critical radius crosses the
transformation value. The final distribution function of membranes is
highly peaked as a result of the very fast early growth.

\section{Conclusions.}
 
We have presented a study of the growth processes for amphiphilic
membranes. These are two dimensional aggregates in a three dimensional 
bath, with peculiar features: the {\it surface vs. volume}
balance which controls the growth for droplets is changed to a 
{\it line vs. surface} balance for planar membranes, or it may be
avoided by closed vesicles, without open edge, at the price of a size
independent curvature energy. Our work here extends the classical
Lifshitz-Slyozov-Wagner theory to consider this problem. We have shown
that, contrary to the nucleation of droplets, there is no
asymptotic limit in which the size distribution of aggregates becomes
independent of the initial conditions. As shown by our numerical
solutions of the coupled equations for planar membranes and spherical
vesicles, the asymptotic form of the distribution function for vesicles
changes dramatically with the initial configuration.  Allowing for the
large uncertainty in the experimental values of several parameters in the
Table I, and for the overall complexity of the problem, our equations 
may provide a first guide to the systematic understanding of these
processes.

It is important to make a comment regarding the absence of fluctuations
in our treatment.
The LSW theory neglects fluctuations because it is expected that
in average the concentration will follow the gradients in chemical 
potential created by the existence of aggregates. But this assumption
conflicts with some other limits taken in the theory in particular large
times, i.e. large domains, and small supersaturations. In this case 
the gradients created by the growth process can be of the same order
or lower than the gradients created by fluctuations in the concentration.
That means that some aggregates could shrink even though their radius
is larger than the critical radius and this effect can change the
size distribution function at late enough times. In particular for vesicles,
this theory predicts a complete degeneracy in the equilibrium size distribution
while fluctuations are expected to break this degeneracy imposing the
the distribution function which maximizes entropy \cite{Morse}.

\acknowledgements

This work was supported by the Direcci\'on General de Investigaci\'on
Cient\'{\i}fica y T\'ecnica (Spain) under grants number PB91-0090  and
PB95-0005.

%%%%%%%%%%%%%%%%%%%%%%%%%%%%%%%%%%%

%%%%%%%%%%%%%%%%%%%%%%%%%%%%%%%%%%%

\begin{table}
\squeezetable
\caption{ Orders of magnitude of some parameters used in the text.  $R_T$ has 
been estimated from the spherical cap model,
and see text Section II.C for the estimation of $\Delta_0$. 
%The estimation
%of $k_s$ and $\lambda$ has been taken  from \cite{ks} and \cite{lambda}
%respectively.
}
\begin{tabular}{lll}
$k_B T$     &  $4 \times 10^{-21}$  &  $J$                     \\
$\rho_v$    &  $10^{18}$            &  $molec/m^3$             \\
$\chi$      &  $2.4\times 10^{38}$  &  $molec^2/J$             \\
$k_s$       &  $10^{-18}$           &  $J$                     \\
$\lambda$   &  $10^{-11}$           &  $J/m$                   \\
$\Gamma$    &  $10^{19}$            &  $molec/m^2$             \\
$R_T$       &  $2\times 10^{-7}$    &  $m$                     \\
$\Delta_0$  &  $3\times 10^{-23}$   &  $T/molec$
\end{tabular}
\end{table}

% FIGURE 1
\begin{figure}
\caption{ Curvature energy of a membrane as a function of its shape for
a {\it spherical cap model} (see text). ${\cal H}(A,R)$ is the curvature energy
(in units of $\kappa_s$) for a membrane of area, $A$, and edge circle of
radius, $R$ (in dimensionless units).
$u=R \left( \pi /A \right)^{1/2}$, where $u=0$ corresponds to spherical 
vesicles and $u=1$ corresponds to planar membranes.  The area for
the spontaneous shape transition is
$A_c=\pi \left( 4 \kappa_s /\lambda \right)^2$}
\end{figure}

% FIGURE 2
\begin{figure}
\caption{ Time evolution of the size distribution function 
in dimensionless units, see eq. (\ref{units}), for initial conditions 
where $90\%$ of 
the excess particles particles are forming 
membranes (see text for details). $x=x_T=66.7$ is the radius for the
spontaneous shape transition from membrane to vesicle. The region $x < x_T$
refers to planar membranes and $x > x_T$ to spherical vesicles.}
\end{figure}

% FIGURE 3
\begin{figure}
\caption{  Critical Radius versus time
in dimensionless units, see eq. (\ref{units}), for initial conditions 
where $90 \%$ of the excess particles particles are forming 
membranes (see text for details). The dotted line corresponds to
$x=x_T$.}
\end{figure}

% FIGURE 4
\begin{figure}
\caption{ Time evolution of the size distribution function 
in dimensionless units, see eq. (\ref{units}), for initial conditions 
where $5 \%$ of the excess particles particles are forming 
membranes (see text for details). $x=x_T=66.7$ is the size of spontaneous
shape transition from membrane to vesicle. The region $x < x_T$
refers to planar membranes and $x > x_T$ to spherical vesicles.}
\end{figure}

% FIGURE 5
\begin{figure}
\caption{  Critical Radius versus time
in dimensionless units, see eq. (\ref{units}), for initial conditions 
where $5 \%$ of the excess particles particles are forming 
membranes (see text for details). The dotted line corresponds to
$x=x_T$.}
\end{figure}


\begin{references}

\bibitem{Lindman85} Lindman B {\it Physics of Amphiphiles: Micelles,
Vesicles and Microemulsions}, Eds. V. Degiorgio and M. Corti, 
(Amsterdam:North-Holland) (1985).

\bibitem{Seddon90} Seddon J M {\it Biochimica et Biophysica Acta}
{\bf 1031} 1 (1990).

\bibitem{Gelbart94} Eds. Gelbart W M, Roux D and Ben-Shaul A 
{\it Micelles, Membranes, Microemulsions and Monolayers.}
(Berlin:Springer) (1994).

\bibitem{Gompper94}  Gompper G and Schick M {\it Self-assembling
amphiphilic systems} in {\it Phase Transitions and Critical Phenomena}
ed. C. Domb and J. Lebowitz, (London:Academic Press) (1994).

\bibitem{Israelachvili85}  Israelachvili J N 
{\it Physics of Amphiphiles: Micelles, Vesicles
and Microemulsions}, Ed. V. Degiorgio and M. Corti, (Amsterdam:North-Holland)
(1985).

\bibitem{Helfrich73} Helfrich W., Z. Naturforsch {\bf 28c}, 693 (1973).
 
\bibitem{Nelson89} Nelson D, Piran T and Weinberg S (eds.) 
{\it Statistical Mechanics of Membranes and Surfaces} (Singapore:World
Scientific) (1989)

\bibitem{Lipowski91} Lipowski R., Nature {\bf 349}, 475 (1991).  

\bibitem{Lipowski92} Lipowski R., Richter D. and Kremer K.
(eds.) {\it The Structure and Conformation of Amphiphilic
Membranes} (Berlin:Springer) (1992).

\bibitem{biomembranes} Gennis R. B. {\it Biomembranes: Molecular 
Structure and Function} (Berlin:Springer) (1989). 

\bibitem{Gunton} Gunton J.D. and Droz M.  {\it Introduction to
the Theory of Metastable and Unstable States} (Berlin:Springer)
(1983). 

\bibitem{Gene} J.D. Watson, N.H. Hopkins, J.W. Robert, J.A. Steitz
and A.M. Weiner, {\it Molecular Biology of the Gene} Benjamin/Cummings
(1989).

\bibitem{LS} I. M. Lifshitz and V. V. Slyozov, J. Phys. Chem. Solids,
{\bf 19}, 35 (1961). 

\bibitem{Wagner} C. Wagner, Z. Elektrochem. {\bf 65}, 581 (1961).

\bibitem{LP} I. M. Lifshitz and L. P. Pitaevskii, {\it Physical
Kinetics}, Landau and Lifshitz Course of Theoretical Physics, vol. 10,
Pergamon (1981).

\bibitem{Arfken} See for example G. Arfken {\it Mathematical Methods for
Physicists}, Third edition (1985), Academic page 661.

\bibitem{numrecipes}  W. H. Press, S. A. Teukolsky, W. T. Vetterling
and B. P. Flannery; {\it Numerical Recipes in Fortran, (second edition)},  
Cambridge University Press, (1992).

\bibitem{Morse}  D. C. Morse and S. T. Milner; {\it Europhys. Lett.},  
{\bf 26}, 565 (1994).

%\bibitem{ks} 


\end{references}
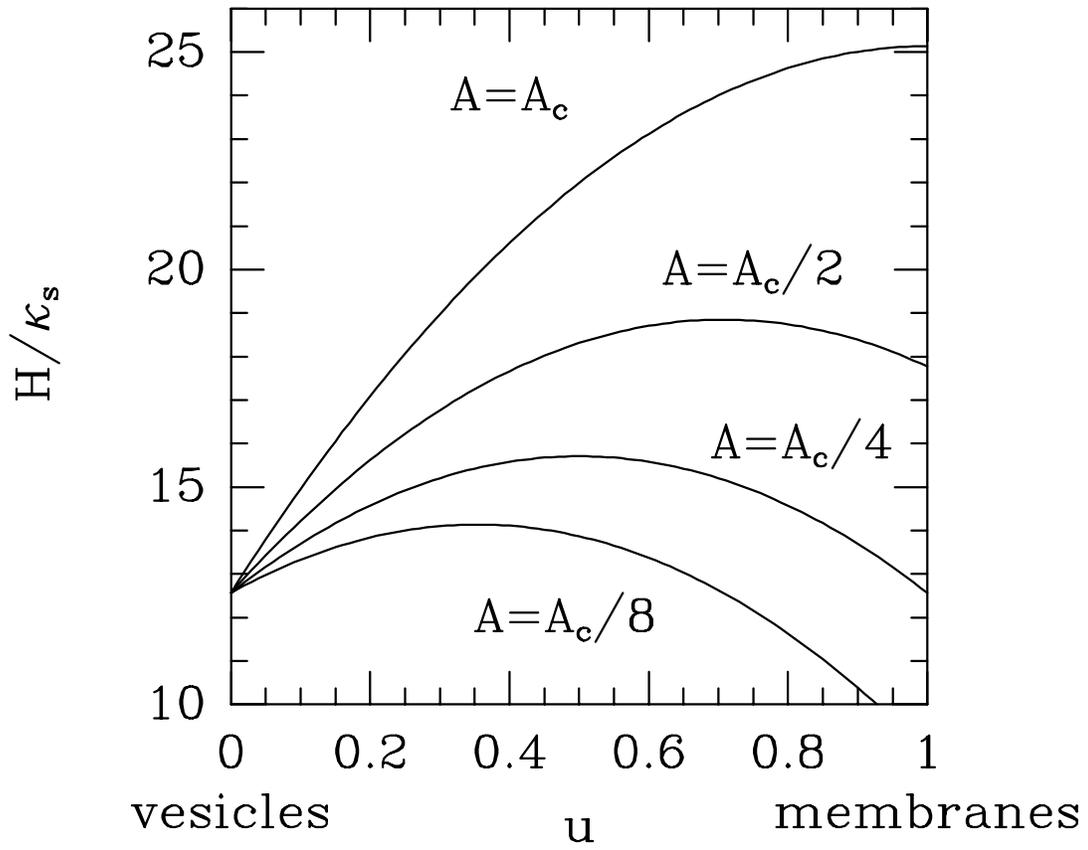
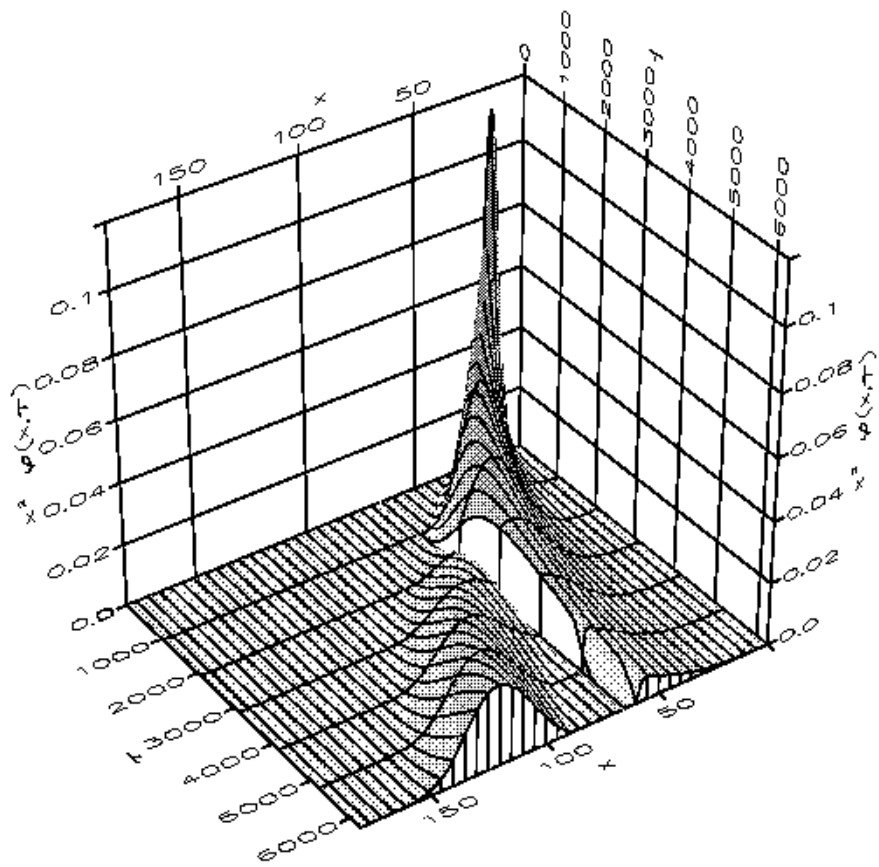
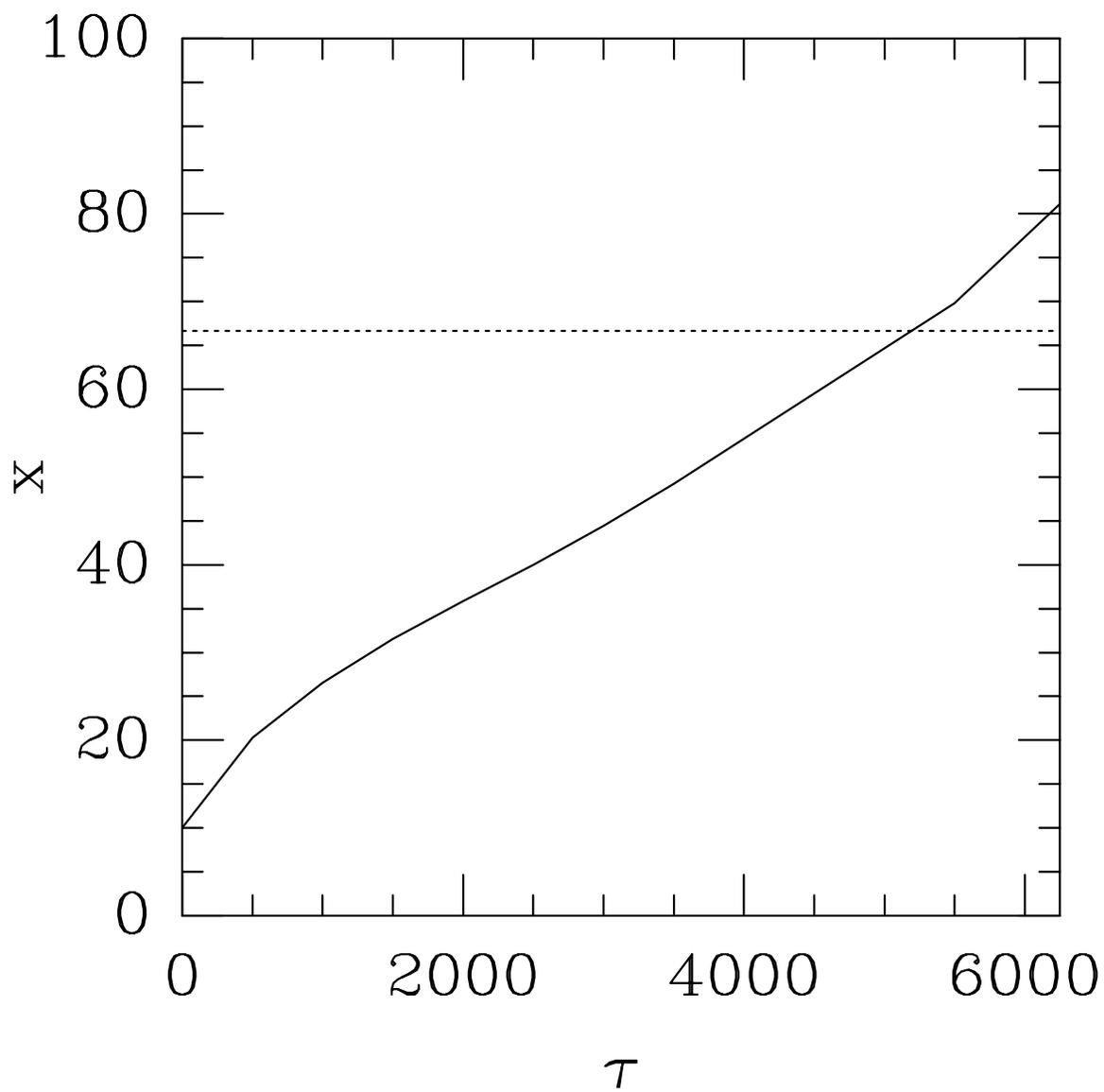
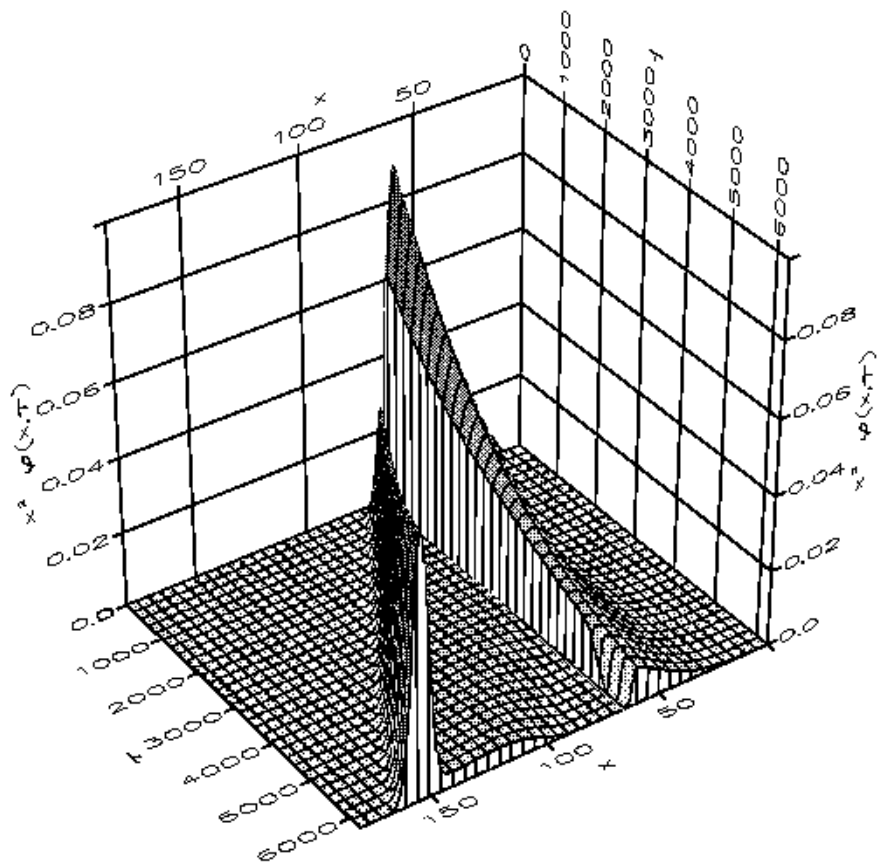
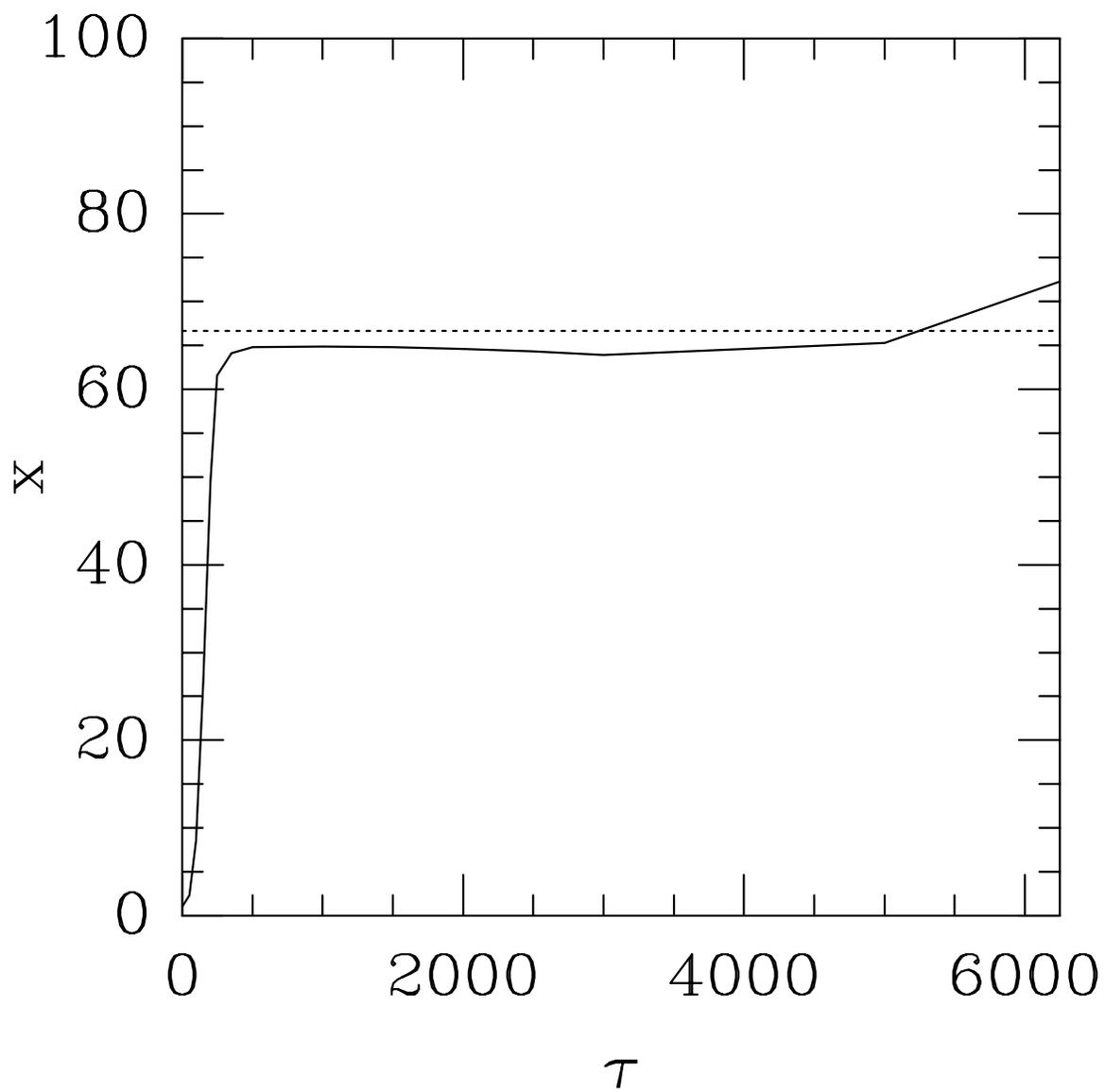
\end{document}